\documentclass[twocolumn,final]{svjour3}

\smartqed

\usepackage{graphicx,amssymb}
\usepackage{epsf,psfig}

\journalname{Nonlinear Dynamics}

\begin{document}

\title{Are semi-numerical methods an effective tool for locating periodic orbits in 3D potentials?}

\author{Euaggelos E. Zotos \and Nicolaos D. Caranicolas}

\institute{Department of Physics, \\
Section of Astrophysics, Astronomy and Mechanics, \\
Aristotle University of Thessaloniki \\
GR-541 24, Thessaloniki, Greece \\
\email{evzotos@astro.auth.gr (Euaggelos E. Zotos)}}

\date{Received: 9 April 2012 / Accepted: 2 May 2012 / Published online: 12 July 2012}

\titlerunning{Are semi-numerical methods an effective tool for locating periodic orbits in 3D potentials?}

\authorrunning{Euaggelos E. Zotos \& Nicolaos D. Caranicolas}

\maketitle

\begin{abstract}

A semi-numerical method is used in order to locate the position and calculate the period of periodic orbits in a 3D composite bisymmetrical potential, in a number of resonant cases. The potential consists of a 3D harmonic oscillator and a Plummer sphere. The outcomes are compared with results found using the numerical integration of the equations of motion and the agreement is very good. This agreement strongly suggests, that semi-numerical methods can be used in order to obtain fast and reliable results regarding the position and period of the periodic orbits in 3D composite potentials with a harmonic oscillator part and different kinds of perturbations. Comparison with other methods of obtaining 3D periodic orbits is discussed.


\end{abstract}

\section{Introduction}

A large number of studies over the last five decades, was devoted to locate the position of periodic orbits in a given dynamical system. The basic reason for doing this, is that near the stable periodic orbits there are similar orbits. On the other hand, orbits starting near the unstable periodic orbits can produce large chaotic regions. On this basis, there is no doubt, that periodic orbits represent the backbone of the whole set of orbits and play an important role in understanding the dynamical behavior of a given potential.

Of particular interest are systems made up of perturbed harmonic oscillators. Systems containing 2D harmonic oscillators were studied by a large number of investigators [1-7, 11, 18, 21, 23].

In a recent paper [8], we presented a semi-numerical method, which was used in order to find the position and the period of periodic orbits in the resonant cases $\omega_1 : \omega_2 = n:m$, in the 2D potential
\begin{equation}
V(x,y) = \frac{1}{2}\left(\omega_1^2 x^2 + \omega_2^2 y^2 \right) - \frac{\mu}{\left(x^2 + y^2 + \alpha^2 \right)^{1/2}}.
\end{equation}

As the results of the above research were in good agreement with the outcomes given by the numerical integration, we decided to extend our semi-numerical method to the 3D potential
\begin{eqnarray}
V(x,y,z) &=& \frac{1}{2}\left(\omega_1^2 x^2 + \omega_2^2 y^2 + \omega_3^2 z^2 \right) \nonumber \\
&-& \frac{\mu}{\left(x^2 + y^2 + z^2 + \alpha^2 \right)^{1/2}},
\end{eqnarray}
where in Eq. (2) $\omega_1$, $\omega_2$ and $\omega_3$ are the unperturbed frequencies of the oscillations along the $x$, $y$ and $z$ axis respectively, while $\mu$ and $\alpha$ are parameters. This potential can be considered to describe local motion in the central parts of a galaxy.

The equations of motion for a test particle with a unit mass are
\begin{eqnarray}
\ddot{x} &=& - \left[\omega_1^2 + \frac{\mu}{\left(x^2 + y^2 + z^2 + \alpha^2 \right)^{3/2}}\right] x
= - w_1^2 x, \nonumber \\
\ddot{y} &=& - \left[\omega_2^2 + \frac{\mu}{\left(x^2 + y^2 + z^2 + \alpha^2 \right)^{3/2}}\right] y
= - w_2^2 y, \nonumber \\
\ddot{z} &=& - \left[\omega_3^2 + \frac{\mu}{\left(x^2 + y^2 + z^2 + \alpha^2 \right)^{3/2}}\right] z
= - w_3^2 z,
\end{eqnarray}
where the dot indicates derivative with respect to the time.

The corresponding Hamiltonian is written
\begin{eqnarray}
H &=& \frac{1}{2}\left(p_x^2 + p_y^2 + p_z^2 + \omega_1^2 x^2 + \omega_2^2 y^2 + \omega_3^2 z^2 \right) \nonumber \\
&-& \frac{\mu}{\left(x^2 + y^2 + z^2 + \alpha^2 \right)^{1/2}} = h,
\end{eqnarray}
where $p_x$, $p_y$ and $p_z$ are the momenta per unit mass conjugate to $x$, $y$ and $z$ respectively, while $h > 0$ is the numerical value of the Hamiltonian, which is conserved. The aim of this work is to find the position of the periodic orbits and the corresponding period in the resonant cases $\omega_1 : \omega_2 : \omega_3 = n:m:l$, where $n$, $m$ and $l$ are positive integers, using a semi-numerical procedure and to compare the results with the corresponding outcomes given by the numerical integration of the equations of motion. In our numerical calculations we shall use the values $\mu = 0.001$ and $\alpha = 0.25$, while the value of the energy $h$ will be treated as a parameter. For the numerical integration of the equations of motion, a Bulirsch-St\"{o}er method in double precision is used. The accuracy of the calculations is checked by the constancy of the energy integral (4), which is conserved up to the fifteenth significant decimal figure.

The paper is organized as follows: In Section 2 we use some theoretical arguments to find the position and period of the 1:1:1 resonant periodic orbits. In the same Section, we present a Table for the periods of the straight-line periodic orbits. Moreover, we make a comparison between the periods of the straight-line orbits derived using the semi-numerical method and the corresponding periods obtained using the numerical integration. In Section 3 we compare the semi-numerical results with the outcomes given by the numerical integration for a number of resonant cases. Finally, a discussion and the conclusions of this work are presented in Section 4.

\section{The 1:1:1 resonance}

In this Section, we shall study the case when $\omega_1 = \omega_2 = \omega_3 = \omega$. In this case the system has a spherical symmetry and potential (2) takes the form
\begin{equation}
V(r) = \frac{1}{2}\omega^2 r^2 - \frac{\mu}{\left(r^2 + \alpha^2 \right)^{1/2}},
\end{equation}
where $r^2 = x^2 + y^2 + z^2$. All the three components $L_x$, $L_y$ and $L_z$ of the test particle's angular momentum and the total angular momentum $L^2 = L_x^2 + L_y^2 + L_z^2$ are conserved. The system has a circular periodic orbit. Using elementary calculations we find that the radius of this circular orbit $r_c$ and the test particle's total angular momentum $L_c$ are connected through the following equation
\begin{equation}
L_c^2 = r_c^3 \left[\frac{dV}{dr}\right]_{r_c},
\end{equation}
while the corresponding value of the energy is
\begin{equation}
h_c = \frac{1}{2}\frac{L_c^2}{r_c^2} + V(r_c).
\end{equation}

As all the three frequencies $w_1$, $w_2$ and $w_3$ in Eqs. (3) are equal the period of the circular orbit is given by the formula
\begin{equation}
T_c = \frac{2 \pi}{\left[\omega^2 + \frac{\mu}{\left(r_c^2 + \alpha^2 \right)^{3/2}}\right]^{1/2}}.
\end{equation}
\begin{table}[ht]
\centering
\caption{A comparison between the period $T_s$ and the period $T_n$ found by the numerical integration of the equations of motion for the 1:1:1 straight line periodic orbits when $\omega_1 = \omega_2 = \omega_3 = 0.4$.}
\begin{tabular}{|c||c|c|}
\hline
$h$ & $T_s$ & $T_n$ \\
\hline \hline
0.035 & 15.4503 & 15.4546 \\
\hline
0.040 & 15.4857 & 15.4866 \\
\hline
0.045 & 15.5138 & 15.5121 \\
\hline
0.050 & 15.5365 & 15.5328 \\
\hline
0.055 & 15.5551 & 15.5499 \\
\hline
0.060 & 15.5706 & 15.5644 \\
\hline
\end{tabular}
\end{table}

Apart from the circular exact periodic orbit the system (5) has an additional type of exact periodic orbit. These orbits are straight lines going through the origin. We observe that for the motion along the lines
\begin{equation}
y = k x, \ \ \ z = \lambda x,
\end{equation}
the equations of motion (3) become identical, while all three frequencies of the oscillations are always equal. Thus, the straight lines (9) are exact 1:1:1 periodic orbits going through the origin. Our numerical experiments show that the period of the  straight line periodic orbits is given by
\begin{equation}
T_s = \frac{2 \pi}{w_s},
\end{equation}
where the frequency $w_s$ is given by the formula
\begin{equation}
w_s = \left[\omega^2 + \frac{0.84 \mu}{\left(r_a^2 + \alpha^2 \right)^{3/2}}\right]^{1/2},
\end{equation}
where $r_a$ is the radius of the circle $V(r)= h$. Note that all the iso-energetic orbits (9) have the same period. A comparison between the period $T_s$ and the period $T_n$ found by the numerical integration of the equations of motion is made in Table 1. We see that the agreement is very good. In order to quantify the agreement between the results obtained from the semi-numerical methods and those derived from the numerical integration, we calculate the error $T_{err} = \left|(T_n - T_s)/T_s\right|$. For the results of the 1:1:1 resonant orbits, presented in Table 1 we see that $T_{err} \leq 0.04\%$.

The authors would like to clarify that since the nature and also the shape of the circular and straight-line periodic orbits is well-known it is therefore unnecessary to provide the corresponding figure 3D plots.

\section{Comparing semi-numerical results to numerical outcomes}

In this Section, we shall give semi-numerical formulas for the position and the period of the periodic orbits in a number of resonant cases. Furthermore, we shall compare the results given by the semi-numerical method with the outcomes derived from the numerical integration of the equations of motion.

We shall start from the resonant case when $\omega_1 : \omega_2 : \omega_3$ = 1:1:2. We look for periodic orbits intersecting the $z$ axis at $z = z_0$, $(x_0 = y_0 = p_{z0} = 0)$, with velocities $p_{x0} = p_{y0}$. Our numerical calculations indicate that the value of $p_{x0}$ is well approximated, if we set
\begin{equation}
p_{x0}^2 = \frac{z_0^2}{2 \omega_1^2}.
\end{equation}
Setting the above initial conditions in the harmonic part of the Hamiltonian (4), that is when $\mu = 0$, and solving the resulting equation for $z_0$, we find
\begin{equation}
z_0 = \left[ \frac{2 h \omega_1^2}{1 + \omega_1^2 \omega_3^2} \right]^{1/2}.
\end{equation}
Note that Eqs. (12) and (13) give a first approximation for the starting point of  the 1:1:2 periodic  orbit.

For a better approximation we proceed as follows: Using numerical experiments we find that a good approximation for the frequency $w_{3s}$ of the 1:1:2 periodic orbits is given by the formula
\begin{equation}
w_{3s} = \left[\omega_3^2 + \frac{0.48 \mu}{\left(z_s^2 + \alpha^2 \right)^{3/2}}\right]^{1/2}.
\end{equation}
Inserting the above value of the frequency $w_{3s}$ in Eq. (12) we find
\begin{equation}
p_{xs}^2 = \frac{z_s^2}{2 w_{1s}^2},
\end{equation}
where now $z_s$ is the new value of the starting point of the 1:1:2 periodic orbit. Inserting the values of $z_s$, $p_{xs} = p_{ys}$, $x = y = p_z = 0$ and $w_{1s} = w_{3s}/2$ in the Hamiltonian (4) we find the following equation
\begin{equation}
\frac{1}{2} \left(w_{1s}^2 + \frac{1}{w_{3s}^2} \right) z_s^2 -
\frac{\mu}{\left(z_s^2 + \alpha^2 \right)^{1/2}} = h.
\end{equation}

Solving Eq. (16) we obtain the value of $z_s$. Remember, that the value of $p_{xs}$ is given by Eq. (15), while the value of $p_{ys}$ can be found from the energy integral. The period of the 1:1:2 periodic orbits is given by
\begin{equation}
T_s = \frac{4 \pi}{w_{3s}}.
\end{equation}
\begin{table}[ht]
\centering
\setlength{\tabcolsep}{4.0pt}
\caption{Starting positions and periods for the 1:1:2 periodic orbits. Subscript $n$ indicates values found by the numerical integration, while subscript $s$ indicates values found using the semi-numerical method. The values of the parameters are: $\omega_1 = \omega_2 = 0.4$ and $\omega_3 = 0.8$.}
\begin{tabular}{|c||c|c|c|c|c|c|}
\hline
$h$ & $z_s$ & $z_n$ & $p_{xs}$ & $p_{xn}$ & $T_s$ & $T_n$ \\
\hline \hline
0.030 & 0.1005 & 0.1002 & 0.1742 & 0.1739 & 15.4153 & 15.4151 \\
\hline
0.032 & 0.1034 & 0.1031 & 0.1792 & 0.1789 & 15.4187 & 15.4183 \\
\hline
0.034 & 0.1062 & 0.1059 & 0.1842 & 0.1840 & 15.4221 & 15.4218 \\
\hline
0.036 & 0.1090 & 0.1088 & 0.1890 & 0.1888 & 15.4286 & 15.4283 \\
\hline
0.038 & 0.1113 & 0.1115 & 0.1936 & 0.1934 & 15.4286 & 15.4283 \\
\hline
0.040 & 0.1142 & 0.1141 & 0.1982 & 0.1981 & 15.4318 & 15.4317 \\
\hline
\end{tabular}
\end{table}

Table 2 gives the starting positions and periods of the 1:1:2 periodic orbits for several values of the energy $h$. Subscript $n$ indicate values found by the numerical integration, while subscript $s$ indicate values found using the Eqs. (15), (16) and (17). The values of the parameters are: $\omega_1 = \omega_2 = 0.4$ and $\omega_3 = 0.8$. One can see, that the agreement is very good. We calculate the errors $z_{err} = \left|(z_n - z_s)/z_s\right|$, $p_{x_{err}} = \left|(p_{xn} - p_{xs})/p_{xs}\right|$ and $T_{err} = \left|(T_n - T_s)/T_s\right|$. For the results of the 1:1:2 resonant orbits, presented in Table 2 we find that $z_{err} \leq 0.30\%$, $p_{x_{err}} \leq 0.18\%$, while $T_{err} \leq 0.03\%$. Figure 1 shows the 1:1:2 periodic orbit, for the above values of the parameters when $h = 0.034$. The initial conditions are: $z_0 = z_n = 0.1059, x_0 = y_0 = p_{z0} = 0, p_{x0} = p_{xn} = 0.1840$, while the value of $p_{y0}$ is found from the energy integral in all cases. The integration time for the 1:1:2 periodic orbit shown in Fig. 1 is $t = T_n = 15.4218$.
\begin{figure}[!tH]
\centering
\resizebox{\hsize}{!}{\rotatebox{0}{\includegraphics*{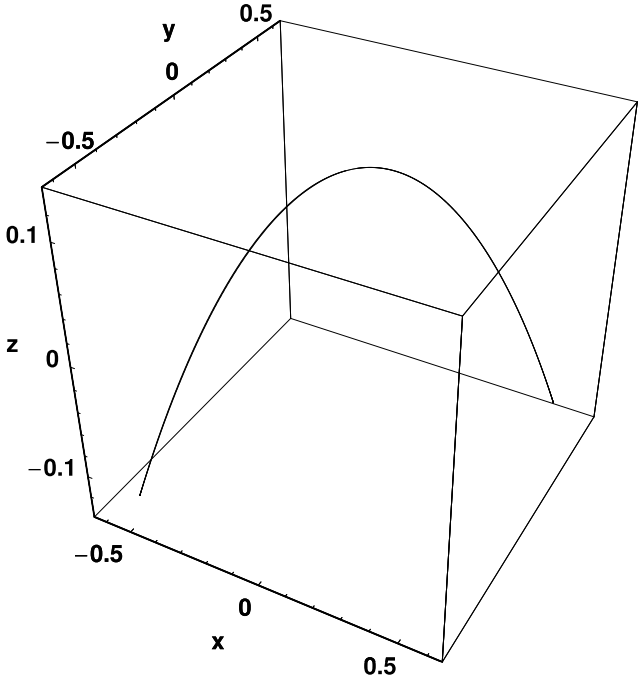}}}
\caption{A 1:1:2 resonant periodic orbit starting from the $z$ axis. The values of parameters are: $\omega_1 = \omega_2 = 0.4, \omega_3 = 0.8$ and $h = 0.034$. The initial conditions are: $z_0 = z_n = 0.1059, x_0 = y_0 = p_{z0} = 0, p_{x0} = p_{xn} = 0.1840$, while the value of $p_{y0}$ is always found from the energy integral. The integration time for this orbits is $t = T_n = 15.4218$.}
\end{figure}

Now we proceed to study the resonant case when $\omega_1 : \omega_2 : \omega_3$ = 2:3:3. We shall study periodic orbits intersecting the $x$ axis at a point $x_s$, $(y_s = z_s = p_{xs} = 0)$ with velocities $p_{zs} = p_{ys}$. In this case, the numerical calculations suggest that a very good approximation for the $x_s$ is given by the formula
\begin{equation}
x_s = \frac{1}{\omega_1}\sqrt{\frac{\omega_2}{\omega_1} \left(h + 0.01 \mu \right)}.
\end{equation}
If we insert the above value of $x_s$ with $y_s = z_s = p_{xs} = 0$ and $p_{zs} = p_{ys}$ in the Hamiltonian (4) we find
\begin{equation}
p_{zs} = \sqrt{h - V_s},
\end{equation}
where
\begin{equation}
V_s = \frac{1}{2} \omega_1^2 x_s^2 - \frac{\mu}{\left(x_s^2 + \alpha^2 \right)^{1/2}}.
\end{equation}
The value of the corresponding period is given by
\begin{equation}
T_s = \frac{4 \pi}{w_{1s}},
\end{equation}
where
\begin{equation}
w_{1s} = \left[\omega_1^2 + \frac{1.5 \mu}{\left(x_s^2 + \alpha^2 \right)^{3/2}}\right]^{1/2}.
\end{equation}
\begin{table}[ht]
\centering
\setlength{\tabcolsep}{4.0pt}
\caption{Similar to Table 2 but for the 2:3:3 resonant periodic orbits. The values of the parameters are: $\omega_1 = \omega_2 = 0.4$ and $\omega_3 = 0.6$.}
\begin{tabular}{|c||c|c|c|c|c|c|}
\hline
$h$ & $x_s$ & $x_n$ & $p_{zs}$ & $p_{zn}$ & $T_s$ & $T_n$ \\
\hline \hline
0.055 & 0.7182 & 0.7181 & 0.1227 & 0.1226 & 31.0863 & 31.0725 \\
\hline
0.060 & 0.7501 & 0.7500 & 0.1275 & 0.1274 & 31.1221 & 31.0970 \\
\hline
0.065 & 0.7807 & 0.7806 & 0.1321 & 0.1322 & 31.1519 & 31.1185 \\
\hline
0.070 & 0.8101 & 0.8100 & 0.1366 & 0.1365 & 31.1771 & 31.1376 \\
\hline
0.075 & 0.8386 & 0.8385 & 0.1410 & 0.1411 & 31.1984 & 31.1546 \\
\hline
0.080 & 0.8661 & 0.8659 & 0.1452 & 0.1453 & 31.2168 & 31.1698 \\
\hline
\end{tabular}
\end{table}
\begin{figure}[!tH]
\centering
\resizebox{\hsize}{!}{\rotatebox{0}{\includegraphics*{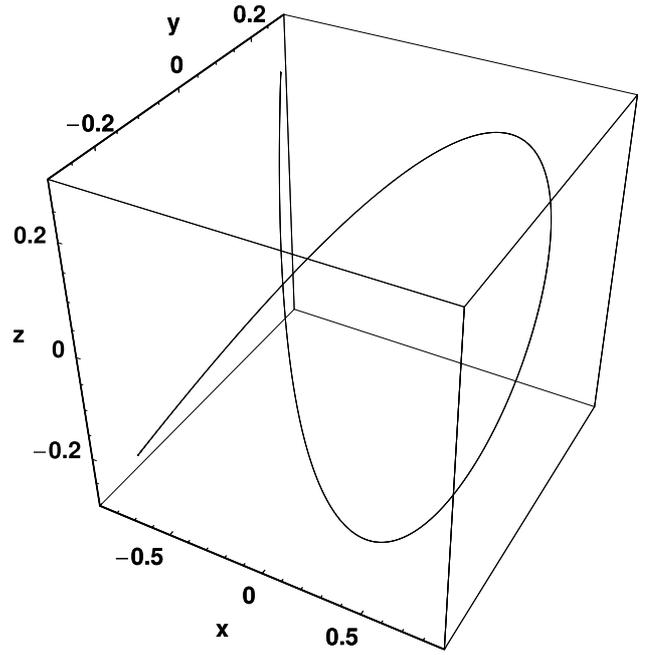}}}
\caption{A 2:3:3 resonant periodic orbit, starting from the $x$ axis when $\omega_1 = 0.4, \omega_2 = \omega_3 = 0.6$ and $h = 0.07$. The initial conditions are: $x_0 = x_n = 0.8100$ with $y_0 = z_0 = p_{x0} =0$ and $p_{z0} = p_{zn} = 0.1365$, while the integration time is $t = T_n = 31.1376$.}
\end{figure}

Table 3 is similar to Table 2 but for the 2:3:3 periodic orbits when $\omega_1 = 0.4$ and $\omega_2 = \omega_3 = 0.6$. As one can see, the agreement between the results given by the numerical integration and the outcomes from the semi-numerical formulas (18), (19) and (17) is very good. For the results of the 2:3:3 resonant orbits, given in Table 3 we find that $x_{err} \leq 0.02\%$, $p_{z_{err}} \leq 0.07\%$, while $T_{err} \leq 0.15\%$. Figure 2 shows the 2:3:3 periodic orbit, for the above values of parameters when $h = 0.07$. The initial conditions are: $x_0 = x_n = 0.8100$ with $y_0 = z_0 = p_{x0} = 0$ and $p_{z0} = p_{zn}= 0.1365$. The integration time for the 2:3:3 periodic orbit shown in Fig. 2 is $t = T_n = 31.1376$. Here we must emphasize that formulae (18), (19) and (20) give very good results also for the resonant cases $\omega_1 : \omega_2 : \omega_3$ = 3:2:2 and $\omega_1 : \omega_2 : \omega_3$ = 4:3:3. The value of the corresponding period $T_s$ for the resonant case 3:2:2 is given by the relation
\begin{equation}
T_s = \frac{6 \pi}{w_{1s}},
\end{equation}
where
\begin{equation}
w_{1s} = \left[\omega_1^2 + \frac{0.52 \mu}{\left(x_s^2 + \alpha^2 \right)^{3/2}}\right]^{1/2}.
\end{equation}

In the resonant case 4:3:3 the period is given by equation
\begin{equation}
T_s = \frac{8 \pi}{w_{1s}},
\end{equation}
where
\begin{equation}
w_{1s} = \left[\omega_1^2 + \frac{0.63 \mu}{\left(x_s^2 + \alpha^2 \right)^{3/2}}\right]^{1/2}.
\end{equation}
\begin{figure}[!tH]
\centering
\resizebox{\hsize}{!}{\rotatebox{0}{\includegraphics*{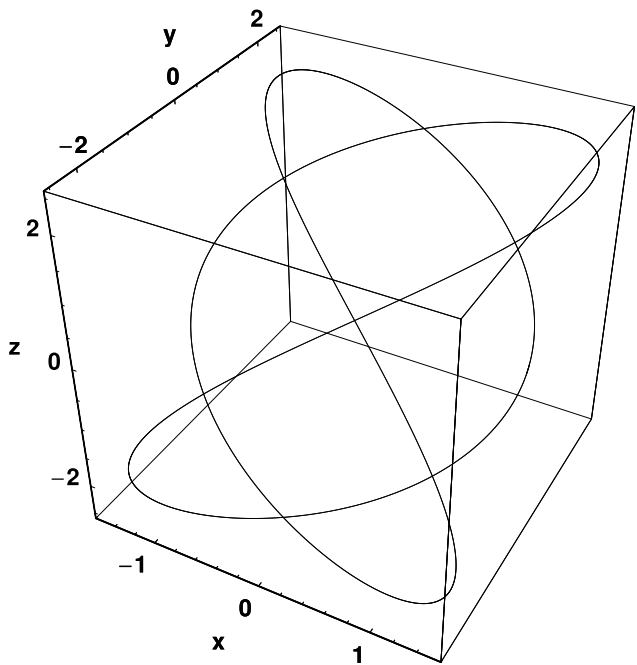}}}
\caption{A 3:2:2 resonant periodic orbit, when $\omega_1 = 0.3$, $\omega_2 = \omega_3 = 0.2$ and $h = 0.30$. The initial conditions are: $x_0 = x_n = 1.4908$ with $y_0 = z_0 = p_{x0} =0$ and $p_{z0} = p_{zn} = 0.4478$, while the integration time is $t = T_n = 62.7798$.}
\end{figure}
\begin{figure}[!tH]
\centering
\resizebox{\hsize}{!}{\rotatebox{0}{\includegraphics*{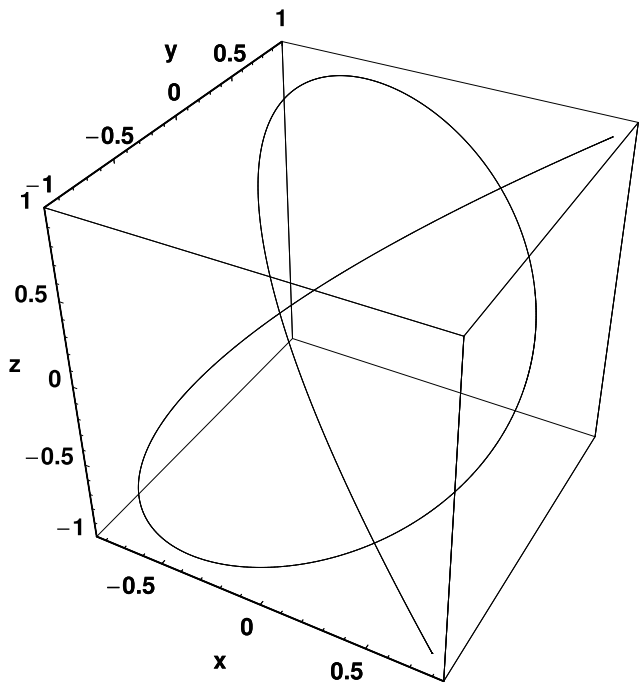}}}
\caption{A 4:3:3 resonant periodic orbit, starting from the $x$ axis, when $\omega_1 = 0.8$, $\omega_2 = \omega_3 = 0.6$ and $h = 0.50$. The initial conditions are: $x_0 = x_n = 0.7654$ with $y_0 = z_0 = p_{x0} =0$ and $p_{z0} = p_{zn} = 0.5600$, while the integration time is $t = T_n = 31.3865$.}
\end{figure}

A comparison between the semi-numerical results and results found by the numerical integration is given in Table 4 for the resonance case 3:2:2 and in Table 5 for the resonance case 4:3:3. The values of the parameters are: $\omega_1 = 0.3, \omega_2 = \omega_3 = 0.2$ for the Table 4 and $\omega_1 = 0.8$, $\omega_2 = \omega_3 = 0.6$ in Table 5. In both cases the agreement is very good. For the results of the 3:2:2 resonant orbits given in Table 4, we find that $x_{err} \leq 0.03\%$, $p_{z_{err}} \leq 0.02\%$, while $T_{err} \leq 0.01\%$, while for the results of the 4:3:3 resonant orbits, presented in Table 5 we see that $x_{err} \leq 0.01\%$, $p_{z_{err}} \leq 0.02\%$, while $T_{err} \leq 0.003\%$. Figure 3 shows the 3:2:2 periodic orbit, for the above values of parameters when $h = 0.30$. The initial conditions are: $x_0 = x_n = 1.4908$ with $y_0 = z_0 = p_{x0} = 0$ and $p_{z0} = p_{zn} = 0.4478$. The integration time for the orbit shown in Fig. 3 is $t = T_n = 62.7798$. The 4:3:3 periodic orbit is given in Figure 4. The values of parameters are as in Table 5 and the value of the energy is $h = 0.50$. The initial conditions are: $x_0 = x_n = 0.7654$ with $y_0 = z_0 = p_{x0} = 0$ and $p_{z0} = p_{zn} = 0.5600$, while the integration time is $t = T_n = 31.3865$.
\begin{table}[ht]
\centering
\setlength{\tabcolsep}{4.0pt}
\caption{Similar to Table 2 but for the 3:2:2 resonant periodic orbits. The values of the parameters are: $\omega_1 = 0.3$ and $\omega_2 = \omega_3 = 0.2.$}
\begin{tabular}{|c||c|c|c|c|c|c|}
\hline
$h$ & $x_s$ & $x_n$ & $p_{zs}$ & $p_{zn}$ & $T_s$ & $T_n$ \\
\hline \hline
0.15 & 1.0541 & 1.0544 & 0.3177 & 0.3176 & 62.6896 & 62.6952 \\
\hline
0.20 & 1.2172 & 1.2174 & 0.3662 & 0.3661 & 62.7375 & 62.7401 \\
\hline
0.25 & 1.3608 & 1.3610 & 0.4091 & 0.4092 & 62.7634 & 62.7647 \\
\hline
0.30 & 1.4907 & 1.4908 & 0.4479 & 0.4478 & 62.7794 & 62.7798 \\
\hline
0.35 & 1.6101 & 1.6102 & 0.4837 & 0.4836 & 62.7899 & 62.7898 \\
\hline
0.40 & 1.7213 & 1.7214 & 0.5169 & 0.5168 & 62.7974 & 62.7973 \\
\hline
\end{tabular}
\end{table}
\begin{table}[ht]
\centering
\setlength{\tabcolsep}{4.0pt}
\caption{Similar to Table 2 but for the 4:3:3 resonant periodic orbits. The values of the parameters are: $\omega_1 = 0.8$ and $\omega_2 = \omega_3 = 0.6.$}
\begin{tabular}{|c||c|c|c|c|c|c|}
\hline
$h$ & $x_s$ & $x_n$ & $p_{zs}$ & $p_{zn}$ & $T_s$ & $T_n$ \\
\hline \hline
0.35 & 0.6404 & 0.6405 & 0.4692 & 0.4693 & 31.3685 & 31.3685 \\
\hline
0.40 & 0.6847 & 0.6848 & 0.5014 & 0.5013 & 31.3761 & 31.3762 \\
\hline
0.45 & 0.7262 & 0.7263 & 0.5315 & 0.5316 & 31.3819 & 31.3820 \\
\hline
0.50 & 0.7655 & 0.7654 & 0.5601 & 0.5600 & 31.3864 & 31.3865 \\
\hline
0.55 & 0.8028 & 0.8027 & 0.5873 & 0.5874 & 31.3900 & 31.3901 \\
\hline
0.60 & 0.8385 & 0.8386 & 0.6133 & 0.6132 & 31.3929 & 31.3930 \\
\hline
\end{tabular}
\end{table}

Finally, we shall consider three additional resonant cases, the cases when $\omega_1 : \omega_2 : \omega_3$ = 5:2:2, $\omega_1 : \omega_2 : \omega_3$ = 5:3:3 and $\omega_1 : \omega_2 : \omega_3$ = 5:4:4. We consider periodic orbits intersecting the $x$ axis at a point $x_s$, $(y_s = z_s = p_{xs} = 0)$ with velocities $p_{zs} = p_{ys}$. In all the above three cases the numerical calculations suggest that a very good approximation for $x_s$ is given by the relation
\begin{equation}
x_s = \frac{1}{\omega_1}\sqrt{\frac{\omega_2}{\omega_1} h}.
\end{equation}
Using the above value of $x_s$ with $y_s = z_s = p_{xs} = 0$, $p_{zs} = p_{ys}$ in the Hamiltonian (4), we find
\begin{equation}
p_{zs} = \sqrt{h - V_s},
\end{equation}
with
\begin{equation}
V_s = \frac{1}{2} \omega_1^2 x_s^2 - \frac{\mu}{\left(x_s^2 + \alpha^2 \right)^{1/2}}.
\end{equation}
\begin{figure}[!tH]
\centering
\resizebox{\hsize}{!}{\rotatebox{0}{\includegraphics*{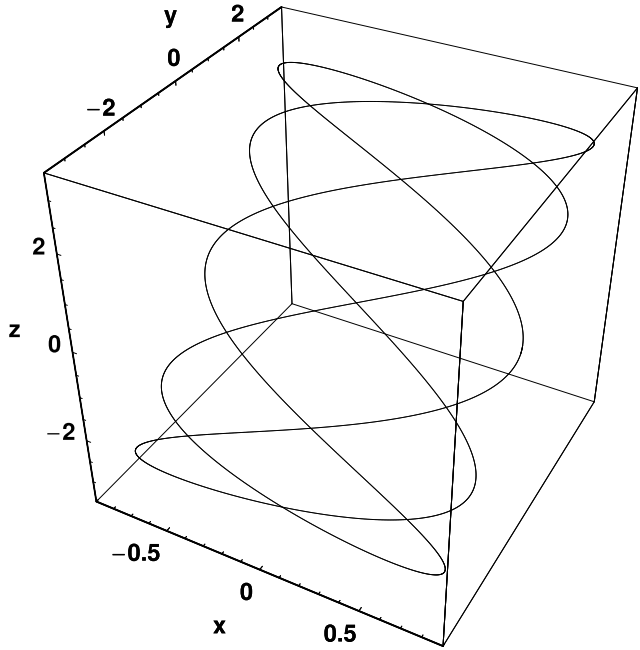}}}
\caption{A 5:2:2 resonant periodic orbit, starting from the $x$ axis, when $\omega_1 = 0.5$, $\omega_2 = \omega_3 = 0.2$ and $h = 0.40$. The initial conditions are: $x_0 = x_n = 0.8001$ with $y_0 = z_0 = p_{x0} = 0$ and $p_{z0} = p_{zn} = 0.5668$, while the integration time is $t = T_n = 62.7944$.}
\end{figure}
\begin{figure}[!tH]
\centering
\resizebox{\hsize}{!}{\rotatebox{0}{\includegraphics*{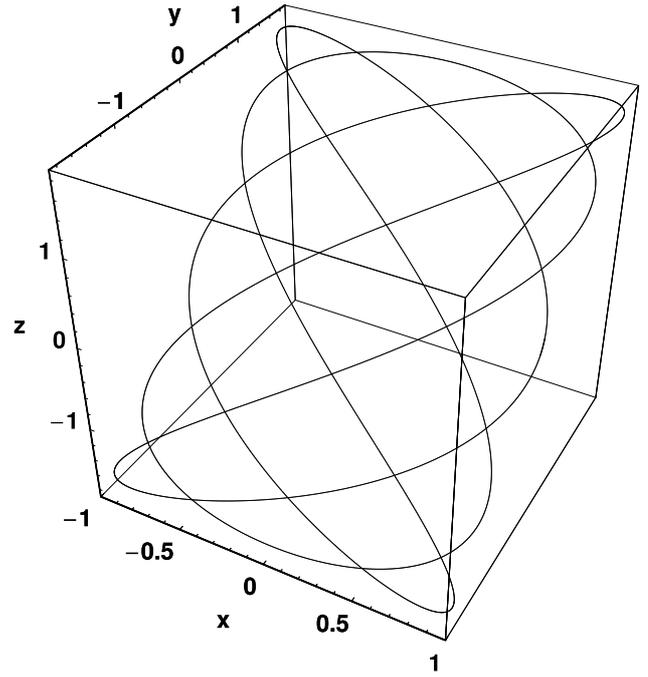}}}
\caption{Similar to Fig. 5 but for the 5:3:3 resonant case. The values of the parameters are: $\omega_1 = 0.5$, $\omega_2 = \omega_3 = 0.3$ and $h = 0.40$. The initial conditions are: $x_0 = x_n = 0.9799$ with $y_0 = z_0 = p_{x0} = 0$ and $p_{z0} = p_{zn} = 0.5300$, while the integration time is $t = T_n = 62.7837$.}
\end{figure}
\begin{figure}[!tH]
\centering
\resizebox{\hsize}{!}{\rotatebox{0}{\includegraphics*{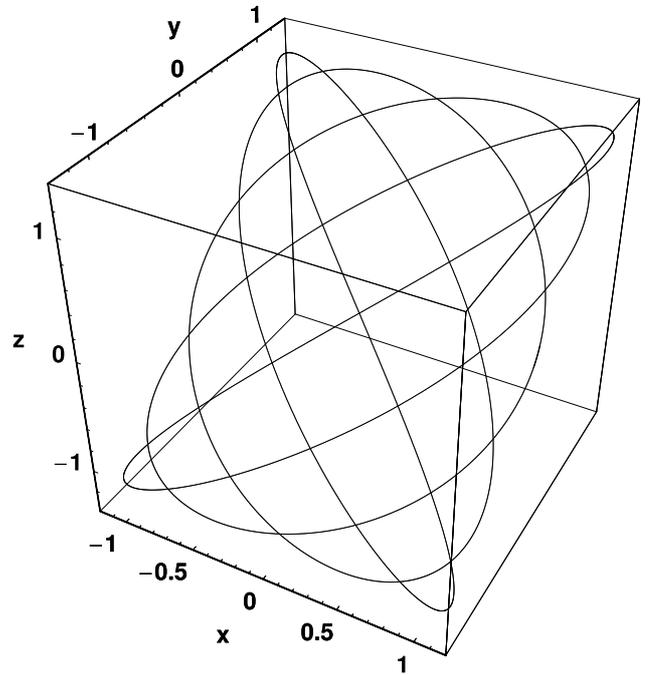}}}
\caption{Similar to Fig. 5 but for the 5:4:4 resonant case. The values of the parameters are: $\omega_1 = 0.5$, $\omega_2 = \omega_3 = 0.4$ and $h = 0.40$. The initial conditions are: $x_0 = x_n = 1.1313$ with $y_0 = z_0 = p_{x0} = 0$ and $p_{z0} = p_{zn} = 0.4907$, while the integration time is $t = T_n = 62.7750$.}
\end{figure}

The value of the corresponding period is given by
\begin{equation}
T_s = \frac{10 \pi}{w_{1s}},
\end{equation}
where
\begin{equation}
w_{1s} = \left[\omega_1^2 + \frac{0.18 \mu}{\left(x_s^2 + \alpha^2 \right)^{3/2}}\right]^{1/2},
\end{equation}
for the 5:2:2 resonance, while
\begin{equation}
w_{1s} = \left[\omega_1^2 + \frac{0.40 \mu}{\left(x_s^2 + \alpha^2 \right)^{3/2}}\right]^{1/2},
\end{equation}
for the 5:3:3 resonance. In the case of the 5:4:4 resonance the value of the frequency $w_{1s}$ is given by the formula
\begin{equation}
w_{1s} = \left[\omega_1^2 + \frac{0.70 \mu}{\left(x_s^2 + \alpha^2 \right)^{3/2}}\right]^{1/2},
\end{equation}
\begin{table}[ht]
\centering
\setlength{\tabcolsep}{4.0pt}
\caption{Similar to Table 2 but for the 5:2:2 resonant periodic orbits. The values of the parameters are: $\omega_1 = 0.5$ and $\omega_2 = \omega_3 = 0.2.$}
\begin{tabular}{|c||c|c|c|c|c|c|}
\hline
$h$ & $x_s$ & $x_n$ & $p_{zs}$ & $p_{zn}$ & $T_s$ & $T_n$ \\
\hline \hline
0.35 & 0.7483 & 0.7484 & 0.5303 & 0.5304 & 62.7858 & 62.7867 \\
\hline
0.40 & 0.8000 & 0.8001 & 0.5667 & 0.5668 & 62.7935 & 62.7944 \\
\hline
0.45 & 0.8485 & 0.8486 & 0.6009 & 0.6010 & 62.7992 & 62.8001 \\
\hline
0.50 & 0.8944 & 0.8945 & 0.6333 & 0.6334 & 62.8036 & 62.8045 \\
\hline
0.55 & 0.9381 & 0.9382 & 0.6641 & 0.6640 & 62.8071 & 62.8097 \\
\hline
0.60 & 0.9798 & 0.9797 & 0.6935 & 0.6934 & 62.8100 & 62.8107 \\
\hline
\end{tabular}
\end{table}
\begin{table}[ht]
\centering
\setlength{\tabcolsep}{4.0pt}
\caption{Similar to Table 2 but for the 5:3:3 resonant periodic orbits. The values of the parameters are: $\omega_1 = 0.5$ and $\omega_2 = \omega_3 = 0.3.$}
\begin{tabular}{|c||c|c|c|c|c|c|}
\hline
$h$ & $x_s$ & $x_n$ & $p_{zs}$ & $p_{zn}$ & $T_s$ & $T_n$ \\
\hline \hline
0.35 & 0.9165 & 0.9166 & 0.4960 & 0.4961 & 62.7733 & 62.7736 \\
\hline
0.40 & 0.9798 & 0.9799 & 0.5301 & 0.5300 & 62.7833 & 62.7837 \\
\hline
0.45 & 1.0392 & 1.0393 & 0.5621 & 0.5622 & 62.7907 & 62.7912 \\
\hline
0.50 & 1.0954 & 1.0955 & 0.5923 & 0.5924 & 62.7964 & 62.7969 \\
\hline
0.55 & 1.1489 & 1.1488 & 0.6212 & 0.6211 & 62.8010 & 62.8014 \\
\hline
0.60 & 1.2000 & 1.2001 & 0.6487 & 0.6488 & 62.8046 & 62.8050 \\
\hline
\end{tabular}
\end{table}
\begin{table}[ht]
\centering
\setlength{\tabcolsep}{4.0pt}
\caption{Similar to Table 2 but for the 5:4:4 resonant periodic orbits. The values of the parameters are: $\omega_1 = 0.5$ and $\omega_2 = \omega_3 = 0.4.$}
\begin{tabular}{|c||c|c|c|c|c|c|}
\hline
$h$ & $x_s$ & $x_n$ & $p_{zs}$ & $p_{zn}$ & $T_s$ & $T_n$ \\
\hline \hline
0.35 & 1.0583 & 1.0582 & 0.4592 & 0.4593 & 62.7636 & 62.7633 \\
\hline
0.40 & 1.1314 & 1.1313 & 0.4908 & 0.4907 & 62.7754 & 62.7750 \\
\hline
0.45 & 1.2000 & 1.2001 & 0.5204 & 0.5203 & 62.7841 & 62.7837 \\
\hline
0.50 & 1.2649 & 1.2650 & 0.5484 & 0.5485 & 62.7909 & 62.7903 \\
\hline
0.55 & 1.3266 & 1.3267 & 0.5751 & 0.5750 & 62.7961 & 62.7956 \\
\hline
0.60 & 1.3856 & 1.3857 & 0.6005 & 0.6006 & 62.8004 & 62.7998 \\
\hline
\end{tabular}
\end{table}

Tables 6, 7 and 8 are similar to Table 5 but for the resonant cases 5:2:2, 5:3:3 and 5:4:4 respectively. The values of the parameters are: $\omega_1 = 0.5, \omega_2 = \omega_3 = 0.2$ for Table 6, $\omega_1 = 0.5, \omega_2 = \omega3 = 0.3$ for Table 7 and $\omega_1 = 0.5, \omega_2 = \omega_3 = 0.4$ for Table 8. We see that, in all cases, the agreement is very good. For the results of the resonant orbits, given in Tables 6, 7 and 8 we find that in all cases $x_{err} \leq 0.01\%$, $p_{z_{err}} \leq 0.02\%$, while $T_{err} \leq 0.001\%$. Figure 5 shows the 5:2:2 periodic orbit, for the above values of the parameters, when $h = 0.40$. The initial conditions are: $x_0 = x_n = 0.8001$ with $y_0 = z_0 = p_{x0} = 0$ and $p_{z0} = p_{zn} = 0.5668$. The integration time for the 5:2:2 periodic orbit shown in Fig. 5 is $t = T_n = 62.7944$. Figure 6 shows the 5:3:3 periodic orbit when $h = 0.40$, while the values of the parameters are as in Table 7. The initial conditions are: $x_0 = x_n = 0.9799$ with $y_0 = z_0 = p_{x0} = 0$ and $p_{z0}= p_{zn} = 0.5300$, while the integration time is $t = T_n = 62.7837$. Finally, a 5:4:4 periodic orbit is shown in Figure 7. The values of the parameters are as in in Table 8, while the value of the energy is $h = 0.40$. The initial conditions are: $x_0 = x_n = 1.1313$ with $y_0 = z_0 = p_{x0} = 0$ and $p_{z0} = p_{zn} = 0.4907$. The integration time for the 5:4:4 periodic orbit shown in Fig. 7 is $t = T_n = 62.7750$.

\section{Discussion and conclusions}

As it was mentioned in Section 1, the periodic orbits play an important role in the dynamical behavior of a Hamiltonian system. In order to locate periodic orbits in 2D and 3D potentials researchers have used a number of numerical or analytical methods [9, 10, 12-17, 22, 20].

In addition to the above methods there are the semi-numerical methods. We have presented semi-numerical results in a number of previous papers [2-5] for polynomial potentials. Semi-numerical methods were also successfully used in celestial mechanics [19]. All the above strongly suggest, that these methods are a very useful tool for the study of the dynamical systems.

In this paper we have studied the periodic motion in a dynamical system composed of a 3D harmonic oscillator and a Plummer potential in several resonant cases $n:m:l$, where $n \leq 5$, $m \leq 4$ and $l \leq 4$. We made this choice for two basic reasons: First, because it is not possible to study all resonant cases in a given potential and second, because the important resonances are those with values of $n$, $m$ and $l$. We considered resonant cases $n:m:l$ in which the periodic orbits are in fact two-dimensional, lying on a particular sub-plane of the three-dimensional space. This issue is very important and essential for our investigation, since it makes easier the construction of the semi-numerical relations regarding the position and the period of the periodic orbit. However, it is in our future plans, to study and provide results about real three-dimensional periodic orbits. Here we must note that potential (2) was also used by the first author to study local motion in the central region of a galaxy [24, 25].

The agreement between the results obtained from the semi-numerical methods and those derived from the numerical integration of the equations of motion depends on the value of the energy $h$. Our numerical experiments indicate, that there is a wide range for the value of the energy in every resonant case, in which the agreement is sufficient enough. However, in every resonant case we studied in the previous sections, we chose six representative values of the energy inside this range, in order to present tables with our data. At this point, it would be useful to refer to how the numerical results indicate or suggest the formulas giving the position and the period of the periodic orbits. In other words, to explain how the formulas are obtained. This was done on a complete empirical basis, using our previous experience, simple expressions, which of course, contain the main parameters entering the Hamiltonian (4) and give outcomes that are close to those given by the numerical integration. If the agreement is not satisfactory, we look for a new formula and so on, until we have achieved a good result.

In particular, the aim of this work was to locate the position and find the period of the periodic orbits in the Hamiltonian (4) in the above mentioned resonant cases using a semi-numerical procedure and compare the results with the outcomes derived from numerical integration. We believe that using semi-numerical methods for finding periodic orbits is very important in 3D Hamiltonian systems, because in these systems it is not easy to find integrals of motion, or to apply analytical methods. The difficulty increases in the cases when we have non polynomial perturbing terms. Furthermore, the comparison of the results given by the semi-numerical method with the results given by numerical integration shows that our method is fast and reliable. On this basis, our semi-numerical method can be considered as an effective and sharp tool in order to find the position and the period of periodic orbits in 3D systems. It is in our plans to try to find more complicated periodic orbits in 3D dynamical systems in the near future.

\section*{Acknowledgements}

The authors would like to express their thanks to the two anonymous referees for the careful reading of the manuscript and their creative suggestions, which have lead to the improvement of the quality and the clarity of the present article.

\end{document}